\documentclass{PoS}
\pdfoutput=1

\title{Towards a Resource-Centric Data Network for Astronomy}

\ShortTitle{Resource-Centric Network for Astronomy}

\author{\speaker{Alberto Accomazzi}\thanks{This work was supported by\
 the Astrophysics Data System project which is funded by NASA grant NNX09AB39G}\\
        Harvard-Smithsonian Center for Astrophysics\\
        E-mail: \email{aaccomazzi@cfa.harvard.edu}}

\author{Michael J. Kurtz\\
        Harvard-Smithsonian Center for Astrophysics\\
        E-mail: \email{mkurtz@cfa.harvard.edu}}

\author{Stephen S. Murray\\
        Harvard-Smithsonian Center for Astrophysics\\
        E-mail: \email{ssm@cfa.harvard.edu}}

\abstract{
Over the past decade, astronomers have been using an increasingly\
larger number of web-based applications and archives to conduct their\
research.  However, despite the early success in creating links across\
projects and data centers, the promise of a single integrated digital\
library environment supporting e-science in astronomy has proven\
elusive.  While some of the issues hampering progress in this area are\
of technical nature, others are rooted in existing policies which\
should be re-analyzed if further rapid progress\
is to be made in this area. This paper describes a proposal that the\
NASA Astrophysics Data System project has put forth in order to\
improve its role as one of the primary discovery portals for astronomers,\
focusing on those aspects which could benefit from an increased level\
of involvement from the community, namely the effort to expose\
astronomy resources as linked data, and the harvesting of\
observational metadata.\
}

\FullConference{Accelerating the Rate of Astronomical Discovery, sps5\\
		August 11-14, 2009\\
		Rio de Janeiro, Brazil}

\begin{document}

\section{Introduction}

Astronomy is an observational science. Our theories of the behavior of
the universe and the objects in it are inspired by observations of
that behavior, and are supported (or rejected) by observations, as
well. Within the past twenty years systems of dense, interlinked data
have become common and expected, fundamentally changing the way people
think and services operate. It is now expected, for example, that from
a listing of a movie at our local theater one can find a synopsis,
reviews, a cast list, a list of all the director's films,
recommendations, and the option to purchase tickets, as well.

Astronomy was one of the first disciplines to benefit from the early
developments of web-based technologies enabling cross-linking of
resources across archives (Accomazzi et al, 1994). Sixteen years ago,
thanks to a collaboration between the 
NASA Astrophysics Data System (ADS) and the 
Centre de Donn{\'e}es Astronomiques de Strasbourg (CDS), it became
possible to go from a list of articles to the abstract of an article
to a list of astronomical objects described in that article to a set
of measurements on one of those objects. Thirteen years ago, again
thanks to a collaboration between the 
ADS and several major data
centers, including NED, HEASARC, MAST, ESO, and Chandra, it became possible
to go from an article abstract to the actual observational data used
to write the article, and then back to all publications describing
each observation.

These connections have enabled astronomers to use the search
capabilities of any of the main archives to locate a dataset or
publication of interest, and then follow the appropriate links to find
related information provided by another archive.  For instance, using
ADS, one could search for papers on X-rays emission in Abell clusters
and filter results to obtain just papers which have links to data
products.  
While this is a very useful way to narrow bibliographic searches,
selecting which of these papers has links to
images or spectra or catalogs still requires a person to click
through all the data links provided in the list of returned papers.
Automating this activity is currently not practical since connections
between one archive and another are purely defined as links between
URLs, and lack semantic and contextual information between the
resources they represent.  In this paper we argue that a tighter
integration of observational and bibliographic metadata will enable the
creation of new connections between resources in astronomy and
applications that will allow users to search, browse, and reason over them.

The existence of these connections between and within
papers and data products represents more that just a convenient
feature to the end-user.  The research process in science today
involves the generation, retrieval, manipulation, analysis, and
publication of digital artifacts which are typically stored in
different archives on the web.  In order to enable scientists to
access these scholarly products, and to document and recreate the
analysis that was carried out on them, a formal description of these
resources and processes is necessary (Pepe et al., 2009).  
Establishing relationships and links among these resources
provides a way to satisfy a number of different goals, among them: 
aggregation (linking together all artifacts used in a study,
including data products, notes, draft papers, software tools);
attribution (properly
acknowledging prior work, tools being used, or datasets being
analyzed); 
preservation (maintaining provenance information and reconstructing 
the workflow used in the research process); and
discovery (following connections between resources may uncover
previously unknown relationships).

\section{A Model for Interlinking Resources}

A model for describing and interlinking resources is provided by the
Linked Data Effort\footnote{http://linkeddata.org}, which is
based on the 
Resource Description Framework (RDF) model (W3C, 2009): a resource has
properties which have values.  For example {\it <PaperA> <isWrittenBy> <AuthorX>}
or {\it <PaperA> <isCitedBy> <PaperB>}.
Chaining these relationships together and performing functions
on the results yields what we call Second Order Operators (Kurtz, 1992).
These operations can have very powerful properties.  For
example, chaining 
{\it <wordPhrase> <isContainedIn> <PapersX>}
with
{\it <PapersX> <areCitedBy> <PapersY>}
yields a list of papers which cite
papers which contain the phrase.  Sorting this list by frequency of
appearance yields review articles on the original word phrase, which
could be any topic such as ``dark energy,'' or ``extrasolar planets.''
The ADS has been internally making use
of the relationships between bibliographic metadata records since its
inception to provide some of its more advanced retrieval capabilities.

An example of what is possible to achieve exploiting these
relationships is provided by the myADS notification service
(Kurtz et al. 2003).  Originally introduced in 2003, this service now 
delivers customized email or RSS updates to over 7,000 users
providing them with updates about the recent technical literature 
in astronomy and physics.  One of the features included in the
update are the most popular and most cited papers on two topics of
interest to the subscriber.  Computation of both lists makes use of
citation and usage relationships data which interlink bibliographic
records in ADS.  Another example of a service that makes use of these
operators is ADS's topic 
search\footnote{http://adsabs.harvard.edu/ads\_abstracts.html}, which can be used
to search the literature for a topic in different ways.
This includes returning a list of the most popular, useful and instructive
papers on a particular subject.  In this context, ``popular,''
``useful,'' and ``instructive'' have been defined in terms of second
order operations on bibliographic, citation, and usage data.

Observational metadata may be similarly modeled: an observation has an
observing proposal, a position in the sky, a time of observation, an
instrument, a telescope, an observer, and a P.I., among other
properties. These properties will likewise have attributes: the
observing proposal will contain words, and may contain a list of
observations; the P.I. may have written papers describing such
observations; the position on the sky may correspond with a known
object; the instrument will have a type (e.g. imager, spectrograph)
and settings (filter, wavelength range, resolution); a combination of
time, position and telescope will yield air mass and moon phase, and
so on.

A brief example will 
illustrate how this knowledge can be used to facilitate a typical researcher's workflow.
Given a list of
the positions of extragalactic globular clusters
one could search for all observations which are IR
spectra. Downloading and analyzing the CaII triplet in these spectra
would allow one to research the metallicity distribution in the nearby
universe. The links from the observing proposals for these spectra to
the journal articles, as well as the proposals themselves, would allow
the user to determine the original intent of the observations, and
whether the proposed work had, indeed, already been done.  This new
system of observation interlinking would clearly be tightly integrated
with the existing infrastructure. For example, the list of
observations of IR spectra of extragalactic globular clusters will
link to a list of observing proposals, which will link to a set of
journal articles which will have been cited by other journal articles,
and so on.

More in general, adopting Linked Data practices to identify, describe,
and connect resources which are already available in our archives 
provides us with a simple way to expose, share, and link pieces of
data, information, and knowledge about them.  One advantage of
this design is that it enables the creation of a global graph, 
i.e. a web of resources which 
reside in different locations but which are connected by semantic
relationships.  The very existence of the graph make it possible for
agents and applications to traverse it in order to describe parts of
it, create new connections within it, or compute new results from it.

\section{Infrastructure and Applications}

While the bibliographic metadata in Astronomy is currently
available from a single  access point (the ADS), the observational
metadata is stored by a number  of archives and projects in different
formats.  Bibliographic searches and Data searches are now being
carried out in separate domains, even though links between the ADS and
the archives allow one to go back and forth between papers and data
products.  However, to date there is no application which combines a
bibliographic search with a ``data'' search based on observational
parameters (such as instrument, wavelength, etc).

An obvious way to enable a more advanced discovery system is to expose
the observational and instrument metadata maintained by the different
archives.  Much of this content already exists, having been created
and curated by groups such as the CDS, ESO, and the NASA archives and data
centers, but the metadata describing this content is currently
unavailable for harvesting in machine-readable format.  Exposing this observational
and technical metadata at the appropriate granularity level would
enable a number of useful applications to be build.  For example,
the ADS project could enhance its holdings by 
harvesting metadata on observational resources which are necessary
to interlink observations and publications more tightly.  Other
organizations would be able to take advantage of the availability of
this metadata as well.  For instance, an institute maintaining a list
of publications by its staff could more easily find out what type of
datasets have been used by them in their research efforts.  A
researcher starting a new project
could identify the core papers on the subject area and find all
datasets studied by them.

A simple prototype interface recently developed by the ADS provides a
good illustration of what types of applications become easily possible
once observational and bibliographic resources become better
interlinked.  The interface in question builds upon the topic search
mentioned in the earlier section and adds to it a view of the 
search results by related astronomical objects.
Figure 1 illustrates an example of the results that are generated from
such an interface.  
Suppose a researcher wants to know what are the extrasolar planets
that are most often referenced in review papers.  Using the topic
search interface, the user can request the list of most
instructive papers on the topic of ``extrasolar planets.''  As results
from the query are retrieved, the system queries the SIMBAD database
to find the list of objects mentioned in each of the papers.  The 
cumulative list
of these object names is then ranked by how frequently each of them
appears in the original set of papers, and is displayed on the left bar.
This list represents a set of so-called ``facets'' which can be used to
further drill into search results, or provide additional views of the
information that is buried within them and otherwise difficult to see.
Using this additional information, a researcher can quickly identify which
objects are most relevant to the topic of interest.  
Applications can also associate one or more actions to each set of
faceted views.
In the example shown in Figure 1, we have provided the option to display
each object via the WorldWideTelescope web client application, but one
could have chosen to use any number of SAMP-enabled applications to
display or map one or more of the objects in question.

It's easy to imagine a variety of similar scenarios in which
bibliographic or observational properties are used to create similar facets.  For
example, when manipulating a list of papers, 
facets could be built on the sets of keywords that are shared amongst
them, or the type of data products associated with them.  While not
all facets may make sense in all cases, there are certainly several
scenarios in which they provide insightful views into the data.  
We hope that in the coming years we will be able to 
investigate the usefulness and impact of faceted views of the
literature when combined with additional observational metadata.

\begin{figure} 
\includegraphics[width=1.0\textwidth]{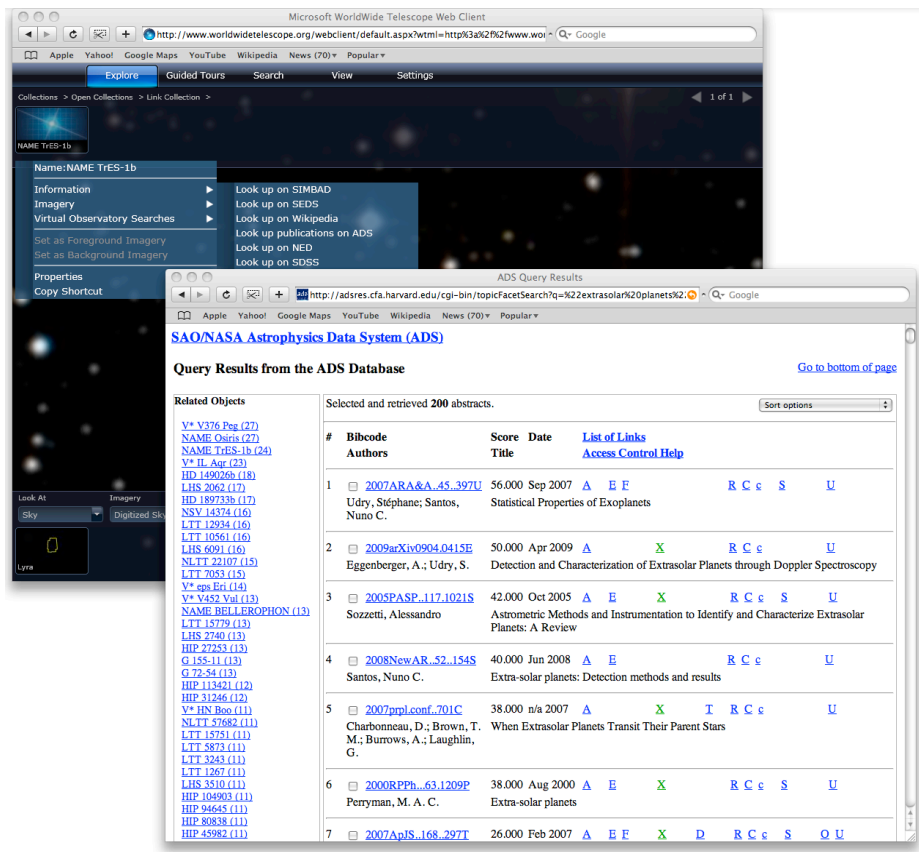} 
\caption{A prototype application implementing object-based facets as a
view into search results.  The original search requested review
articles about ``extrasolar planets.''  From the list of most relevant
objects, the third entry from the top was selected (TrES-1b), and the
object was displayed via a link to the WWT web client.} 
\label{fig1} 
\end{figure}

\section{Discussion}

In this coming era of data intensive science, it is increasingly
important to be able to seamlessly move between scientific results,
the data used to publish them, and the processes used to produce
them.  In addition, scientific research 
requires that we are able to establish the
provenance of data sources and processes operating on this data, so
that research may be repeated, variations on the research be carried
out, and new research on existing datasets be enabled.

While the Virtual Observatory has provided us with the standards and
protocols needed to model and exchange astronomical datasets, the
problem of information discovery in our data-centric world still looms
large.  The key to creating such a system is having access to the
metadata describing the datasets at the appropriate granularity level
so that the proper resources can be identified, described, and 
connections between them can be made.  The technology which can be
used to expose and link these metadata has now matured and is the
foundation of the 
semantic web framework and in particular in the linked data model.
These technologies promise to become widely-adopted standards enabling the 
creation and growth of a ``web of data'' in which machine-readable
content parallels the human-readable content available today.

We believe that the astronomical community would benefit from
embracing this philosophy and adopting these technologies.
In particular, three sets of metadata resources currently 
stored in our archives should be made available 
machine-readable format for harvesting and indexing: observing
proposals, their attributes and links to observations; observations
and their attributes; instruments, their attributes and
capabilities. From a technical standpoint this metadata
should be exposed as RDF and should make use of community-developed
(lightweight) ontologies.  This will allow the use of linked-data
principles to properly connect these resources.  
However, even before these technical aspects are worked
out, we believe that the availability of the metadata to organizations
such as the ADS, the CDS, NED, and other IVOA members will enable the
creation of knowledge bases and applications demonstrating the value of
such an interlinked data system.
Therefore we urge the community to support this effort by recommending
that observatories and archives make available
and permit the harvesting and indexing of metadata describing
publicly accessible datasets, including observing proposals,
observing logs, and FITS headers.

\end{document}